\documentclass[letterpaper,twocolumn,prl,aps,superscriptaddress,amsmath,amssymb,floatfix]{revtex4-1}
\usepackage[latin9]{inputenc}
\usepackage{color}
\usepackage{verbatim}
\usepackage{float}
\usepackage{amsmath}
\usepackage{amssymb}
\usepackage{graphicx}
\usepackage{epstopdf}
\usepackage{esint}
\usepackage[unicode=true,
 bookmarks=true,bookmarksnumbered=false,bookmarksopen=false,
 breaklinks=false,pdfborder={0 0 1},backref=false,colorlinks=true]
 {hyperref}
\hypersetup{
 linkcolor=magenta,urlcolor=blue,citecolor=blue,pdfstartview={FitH},hyperfootnotes=false}

\makeatletter

\usepackage{textcomp}
\usepackage{epstopdf}

\pdfpageheight\paperheight
\pdfpagewidth\paperwidth

\@ifundefined{textcolor}{}{%
 \definecolor{BLACK}{gray}{0}
 \definecolor{WHITE}{gray}{1}
 \definecolor{RED}{rgb}{1,0,0}
 \definecolor{GREEN}{rgb}{0,1,0}
 \definecolor{BLUE}{rgb}{0,0,1}
 \definecolor{CYAN}{cmyk}{1,0,0,0}
 \definecolor{MAGENTA}{cmyk}{0,1,0,0}
 \definecolor{YELLOW}{cmyk}{0,0,1,0}
}

\usepackage{xcolor}
\usepackage{soul}
\definecolor{blue}{rgb}{0,0,1}
\definecolor{red}{rgb}{1,0,0}
\definecolor{green}{rgb}{0,1,0}

\usepackage{soul}

\makeatother

\begin{document}

\address{CAS Key Laboratory of Quantum Information, University of Science
and Technology of China, Hefei, Anhui 230026, China.}
\address{Center for Quantum Information, Institute for Interdisciplinary
Information Sciences, Tsinghua University, Beijing 100084, China}
\address{School of Civil Engineering, Hefei University of Technology,
Hefei 230009, P.R. China.}
\address{CAS Center For Excellence in Quantum Information and Quantum
Physics, University of Science and Technology of China, Hefei, Anhui
230026, China}

\title{High-frequency traveling-wave phononic cavity with sub-micron wavelength }

\author{Xin-Biao Xu}
\address{CAS Key Laboratory of Quantum Information, University of Science
and Technology of China, Hefei, Anhui 230026, China.}
\address{CAS Center For Excellence in Quantum Information and Quantum
Physics, University of Science and Technology of China, Hefei, Anhui
230026, China}

\author{Jia-Qi Wang}
\address{CAS Key Laboratory of Quantum Information, University of Science
and Technology of China, Hefei, Anhui 230026, China.}
\address{CAS Center For Excellence in Quantum Information and Quantum
Physics, University of Science and Technology of China, Hefei, Anhui
230026, China}

\author{Yuan-Hao Yang}
\address{CAS Key Laboratory of Quantum Information, University of Science
and Technology of China, Hefei, Anhui 230026, China.}
\address{CAS Center For Excellence in Quantum Information and Quantum
Physics, University of Science and Technology of China, Hefei, Anhui
230026, China}

\author{Weiting Wang}
\address{Center for Quantum Information, Institute for Interdisciplinary
Information Sciences, Tsinghua University, Beijing 100084, China}

\author{Yan-Lei Zhang}
\address{CAS Key Laboratory of Quantum Information, University of Science
and Technology of China, Hefei, Anhui 230026, China.}
\address{CAS Center For Excellence in Quantum Information and Quantum
Physics, University of Science and Technology of China, Hefei, Anhui
230026, China}

\author{Bao-Zhen Wang}
\address{School of Civil Engineering, Hefei University of Technology,
Hefei 230009, P.R. China.}

\author{Chun-Hua Dong}
\address{CAS Key Laboratory of Quantum Information, University of Science
and Technology of China, Hefei, Anhui 230026, China.}
\address{CAS Center For Excellence in Quantum Information and Quantum
Physics, University of Science and Technology of China, Hefei, Anhui
230026, China}

\author{Luyan Sun}
\address{Center for Quantum Information, Institute for Interdisciplinary
Information Sciences, Tsinghua University, Beijing 100084, China}

\author{Guang-Can Guo}
\address{CAS Key Laboratory of Quantum Information, University of Science
and Technology of China, Hefei, Anhui 230026, China.}
\address{CAS Center For Excellence in Quantum Information and Quantum
Physics, University of Science and Technology of China, Hefei, Anhui
230026, China}

\author{Chang-Ling Zou}
\email{clzou321@ustc.edu.cn}
\address{CAS Key Laboratory of Quantum Information, University of Science
and Technology of China, Hefei, Anhui 230026, China.}
\address{CAS Center For Excellence in Quantum Information and Quantum
Physics, University of Science and Technology of China, Hefei, Anhui
230026, China}
\address{National Laboratory of Solid State Microstructures, Nanjing University, Nanjing 210093, China.}

\begin{abstract}
Thin-film gallium nitride (GaN) as a proven piezoelectric material is a promising platform for the phononic integrated circuits, which hold great potential for scalable information processing processors. Here, an unsuspended traveling phononic resonator based on high-acoustic-index-contrast mechanism is realized in GaN-on-Sapphire with a frequency up to 5 GHz, which matches the typical superconducting qubit frequency. A fivefold increment in quality factor was found when temperature decreases from room temperature ($Q=5000$) to $7\,\mathrm{K}$ ($Q=30000$) and thus a frequency-quality factor product of  $1.5\times10^{14}$ is obtained. Higher quality factors are available when the fabrication process is further optimized. Our system shows great potential in hybrid quantum devices via circuit quantum acoustodynamics.
\end{abstract}
\maketitle

\section{Introduction}

Phonons have been widely used in sensing~\citep{Laenge2008,Liu2016}
and classical communications~\citep{Campbell1998}. For quantum information
processing, phonons have also received significant attention because
of their particular advantages compared to photons: phononic devices
are widely accepted as a universal quantum transducer to connect various quantum systems such as, spin electrons~\citep{Barnes2000,Hermelin2011}, quantum dots~\citep{Naber2006,Gell2008}, NV center~\citep{Golter2016,Golter2016a} and superconducting qubits~\citep{Gustafsson2014,Chu2017,Satzinger2018,Mirhosseini2020}.
Besides, phonons show ultralong lifetime compared to photons. The lifetime
of a phononic mode up to 1.5 seconds has been demonstrated recently
by carefully engineering the energy damping of the integrated phononic
resonator, which makes phononic devices a good candidate for quantum
memory~\citep{MacCabe2020}. As the propagation speed of phonons
is five orders of magnitude smaller than that of microwave photons, the
wavelength of the acoustic wave is also five orders of magnitude smaller
than that of microwaves at the same frequency. By replacing the microwave
devices in circuit quantum electrodynamics (QED) system with integrated
phononic devices, the novel hybrid quantum systems show higher level
of integration and scalability. Furthermore, the interaction between
the superconducting artificial atoms and the GHz phonons, which is
the so-called circuit quantum acoustodynamics (QAD), can explore novel
and diverse dynamic processing~\citep{Wallraff2004,Manenti2017,Guo2017,Andersson2019}.
Although different mechanical systems, such as suspended mechanical
beams~\citep{Lahaye2009}, bulk acoustic wave resonators~\citep{Chu2017}
and phononic-crystal-based acoustic wave resonators~\citep{Mirhosseini2020}
have been proposed and extensively studied for quantum information
processing in the hybrid quantum systems, most of the phononic modes
are localized or traveling without lateral confinement in the 2D phononic
film~\citep{Gustafsson2014,Satzinger2018,Bienfait2019}.

Different from the above mechanical systems, an integrated and unsuspended
phononic integrated circuit (PnIC) platform with traveling phonons
has been recently proposed~\citep{Fu2019,Wang2020,Mayor2021,Shao2021}.
Similar to the optical waveguide, high-acoustic-index-contrast between
the core waveguide and the substrate material allows the phonon waveguide
to support discrete and confined phononic modes. The wavelength of the phonon waveguide ensures a small mode volume, which is essential to realize the strong coupling with other systems. Taking
advantage of advanced nano-fabrication technologies, PnIC with different
functional phononic devices can be constructed. Among them, high frequency
integrated micro-resonator with high quality factor ($Q$) is one
of the important devices~\citep{Moores2018,Shen2020} as it promises
large $Q/V$, which indicates a stronger phonon-matter interaction.
Besides, higher frequency phononic modes also exhibit fewer thermal
phonons at millikelvin temperatures that is critical for quantum coherence
maintenance.

As a mature material,
gallium nitride (GaN) shows good piezoelectric properties and great
potential in PnIC~\citep{RaisZadeh2014}. In this work, a traveling phononic microring resonator with frequencies
up to $5\,\mathrm{GHz}$ with sub-micro acoustic wavelength is demonstrated
on the GaN-on-sapphire (GNOS) platform~\citep{Fu2019}. The mechanical
$Q$-factor of the cavity increases from 5000 to 30000 when the temperature
decreases from room temperature to $7\,\mathrm{K}$.

\begin{figure}
\begin{centering}
\includegraphics[width=1\columnwidth]{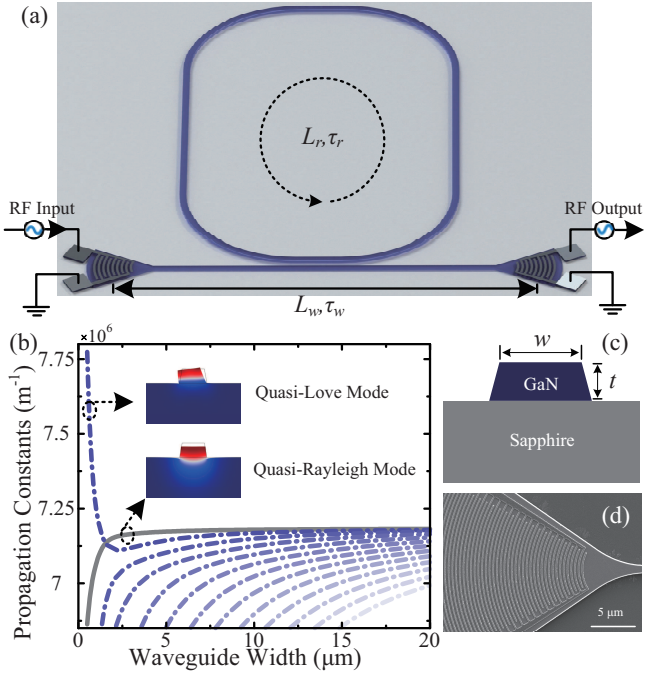}
\par\end{centering}
\caption{\label{fig:1}(a) Schematic diagram of the integrated traveling wave
phononic cavity. $L_{r}$ and $L_{w}$ are the total lengths of the
racetrack resonator and the distance between two IDTs, respectively. $\tau_{r}$
and $\tau_{w}$ are the corresponding traveling times of phonon in
cavity and waveguide with $\tau=L/v_{g}$, where $v_{g}$ is the group
velocity of the phonon. (b) The propagation constant of different
phononic modes along $[1\bar{1}00]$ direction versus the waveguide
width when the phonon frequency is $5\,\mathrm{GHz}$. The gray solid
line represents the quasi-Rayleigh mode. The inset shows both the quasi-Love and Rayleigh modes. (c) The cross-section of the
GaN waveguide on sapphire. (d) The SEM image of IDT
structure.}
\end{figure}

As shown in Fig.~\ref{fig:1}(a), we focus on a simple PnIC configuration because it consists of all the essential passive ingredients for PnIC, including the efficient energy transduction, multiport couplers, and low-loss phonon guiding and confinement.
The phononic mode in the microresonator is excited by the input fan-shaped interdigital transducer (IDT) as Fig.~\ref{fig:1}(d) shows, which converts the RF
signal to the vibration of the phonon waveguide by the piezoelectric effect~\citep{Fu2019}. The inverse process occurs at the output IDT, where the energy of the phonon is converted to the output RF signal. The coupling between the bus waveguide and the cavity relies
on the evanescent coupling.  To match the frequencies of superconducting transmon qubits as well as the wavelength of optical photons for future phonon-based hybridized quantum circuits, we target at the working frequency of $5\,\mathrm{GHz}$. Unlike the previously demonstrated unsuspended phononic circuit at lower frequencies, i.e.
about $200\,\mathrm{MHz}$ in Ref.~\citep{Fu2019} and $3.5\,\mathrm{GHz}$
in Ref.~\citep{Mayor2021}, our devices require the structure dimension to
be less than $1\,\mathrm{\mu m}$ with the waveguide width of $w=500\,\mathrm{nm}$ and the corresponding acoustic wavelength is $916\,\mathrm{nm}$. To enable the evanescent-field coupling between waveguides, we choose a thickness of the GaN film as $t=320\,\mathrm{nm}$.

Properties of the phononic waveguide on the GaN-on-Sapphire system is numerically studied by the finite-element method (COMSOL Multiphysics). The displacement field distribution of the fundamental quasi-Love and quasi-Rayleigh modes are illustrated by the insets of Fig.~\ref{fig:1}(b), with the corresponding
cross-section of the waveguide being shown in Fig.~\ref{fig:1}(c). We
choose the quasi-Rayleigh mode because it is dominated by the out-of-plane
displacement and is more preferable for an efficient excitation by the
IDT in GNOS. Also, the quasi-Rayleigh mode shows considerable evanescent
field components in the substrate with a penetration depth of about
$150\,\mathrm{nm}$, implying efficient coupling between the waveguide
and the ring resonator by setting a gap between them as $g=300$$\,\mathrm{nm}$.
In Fig.~\ref{fig:1}(b), the propagation constant of different phononic
modes at $5\,\mathrm{GHz}$ along $[1\bar{1}00]$ direction of sapphire
as well as the direction of input waveguide is shown by varying the
waveguide width $w$ from $500\,\mathrm{nm}$ to $20\,\mathrm{\mu m}$
corresponding to the waveguide width of the fan-shaped waveguide for
IDT. It shows that the propagation constants of both quasi-Love and
quasi-Rayleigh modes are insensitive to the width of the waveguide
for $w>2\,\mathrm{\mu m}$, thus requiring a period of the IDT finger
of $873\,\mathrm{nm}$ to match the propagation constant of the quasi-Rayleigh
mode in the fan-shaped IDT region.

For the device fabrication, the phononic structures are patterned on the hydrogen
silsesquioxane (HSQ, FOx-16) resist by an electron-beam lithography (EBL).
The pattern is then transferred to the GaN film layer by an optimized
full-etched inductively coupled plasma (ICP) dry etching processing.
A second EBL is implemented to define the IDTs pattern using polymethyl
methacrylate (PMMA) resist. 10-nm-thick titanium and 50-nm-thick aluminum
are deposited followed by a subsequent lift-off process in acetone
to realize the IDTs structure. Fig.~\ref{fig:1}(d) shows the
device images of the IDT. The total length of the fabricated racetrack resonator
is $L_{r}=724\,\mathrm{\mu m}$ with a bending radius $R=50\,\mathrm{\mu m}$.
$\tau_{r}=L_{r}/v_{g,r}$ is the round-trip time,
where $v_{g,r}$ is the average group velocity of the phonon traveling
in the cavity. $L_{w}\approx136\,\mathrm{\mu m}$ and $\tau_{w}$
are the distance between two IDTs and the corresponding traveling
time of phonons with $\tau_{w}=L_{w}/v_{g,w}$, where $v_{g,w}$ is
group velocity of the phonon traveling in the waveguide.

\begin{figure}
\begin{centering}
\includegraphics[width=1\columnwidth]{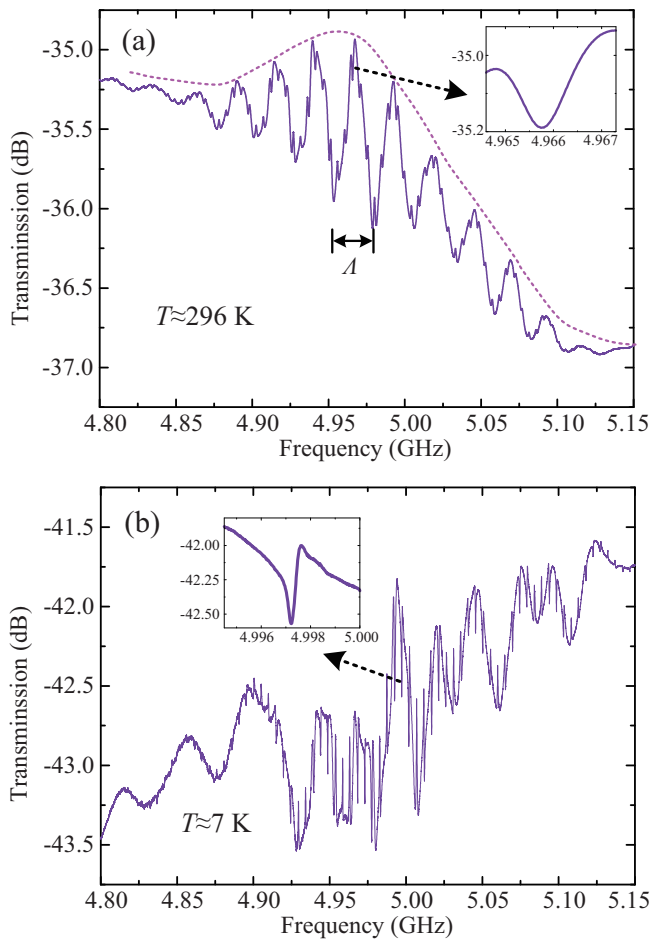}
\par\end{centering}
\caption{\label{fig:2}(a) and (b) are the transmissions of the phonon cavity
at room temperature (296 K) and low temperature (7~K), respectively. The insets are
one of the enlarged resonances of the phononic cavity.}
\end{figure}

The $5\,\mathrm{GHz}$ phononic circuit is characterized by measuring
the transmission spectra of the system by a vector network analyzer
(VNA) with the measurement setup shown in Fig.~\ref{fig:1}(a). The
transmission spectra at room temperature and  $T\approx7\,\mathrm{K}$
are plotted in Fig.~\ref{fig:2}. First of all, we could recognize
an envelope of the spectrum modulation for the results at room temperature,
as shown by the dashed pink line in Fig.~\ref{fig:2}(a). Such an envelope
shows a limited IDT response frequency bandwidth in practice, and
the bandwidth is inversely proportional to the number of IDT periods
(21 for our devices)~\citep{Smith1969}. The extracted full width at half maximum and the center frequency of the IDT are about $150\,\mathrm{MHz}$
and $4.97\,\mathrm{GHz}$, respectively, which matches well with the
theoretical expectation of $160\,\mathrm{MHz}$ and $5\,\mathrm{GHz}$.
The modulation of the envelope resembles the sinusoidal interference
fringes, with a period of about $\varLambda=25.9\,\mathrm{MHz}$.
There are two conjectures on the explanation of the interference:
(1) the Mach-Zendel interferometer (MZI) mechanism with the interference
between the direct phononic waveguide channel and the IDT cross-talk
via air; (2) the phononic Fabry-Perot cavity effect in the straight
waveguide due to the phonon reflections at two IDTs. For the two conjectures,
the difference of the path lengths is $L_{w}$ and $2L_{w}$, respectively.
For the observed $\varLambda$, the two conjectures predict the group
velocities of the phononic waveguide as $v_{g,w}=L_{w}\cdot\varLambda\approx3522.4\,\mathrm{m/s}$
and $v_{g,w}=2L_{w}\cdot\varLambda\approx7044.8\,\mathrm{m/s}$, respectively.

For the resonator modes that we are interested in, we find many shallow
dips or peaks on top of the background envelope. The inset shows the
magnified spectrum of one selected dip, as indicated by the dashed
arrow, corresponding to a $Q$-factor of a few thousands. For these
shallow and sharp resonances, we can roughly estimate the FSR of the
phononic racetrack cavity as $\mathrm{FSR}/2\pi=4.8\,\mathrm{MHz}$, corresponding
to a group velocity of the acoustic wave in the cavity as $\bar{v}_{g,r}=3476\,\mathrm{m/s}$.
Comparing the group velocity to the results of $v_{g,w}$, we confirm
that the modulation envelope of the spectrum with a period of $\varLambda$
is attributed to the MZI mechanism.

The propagation loss of the phonon waveguide is determined by the
geometry of the waveguide, the phononic properties of the material,
and the nano-fabrication processes. Among those, the phononic properties
of the material intrinsically limit the $Q$-factor of the cavity,
which can be improved by upgrading the material growth technologies.
Additionally, the intrinsic phonon loss of the material can also be
reduced at lower temperatures~\citep{Fu2019,Mayor2021}, and thus a
higher $Q$-factor is expected. In Fig.~\ref{fig:2}(b), the transmission
of the same device is measured at $T\approx7\,\mathrm{K}$. It shows
that the extinction ratio of the measured resonances is improved with
the bandwidth decreasing obviously, implying a significant improvement
of the $Q$-factor. The inset presents an enlarged resonance, which
shows a Fano-type resonance~\cite{Limonov2017Sep} and is attributed to the interference
between the phononic narrow resonances and the broadband modulation due
to the MZI mechanism.

\begin{figure}
\begin{centering}
\includegraphics[width=1\columnwidth]{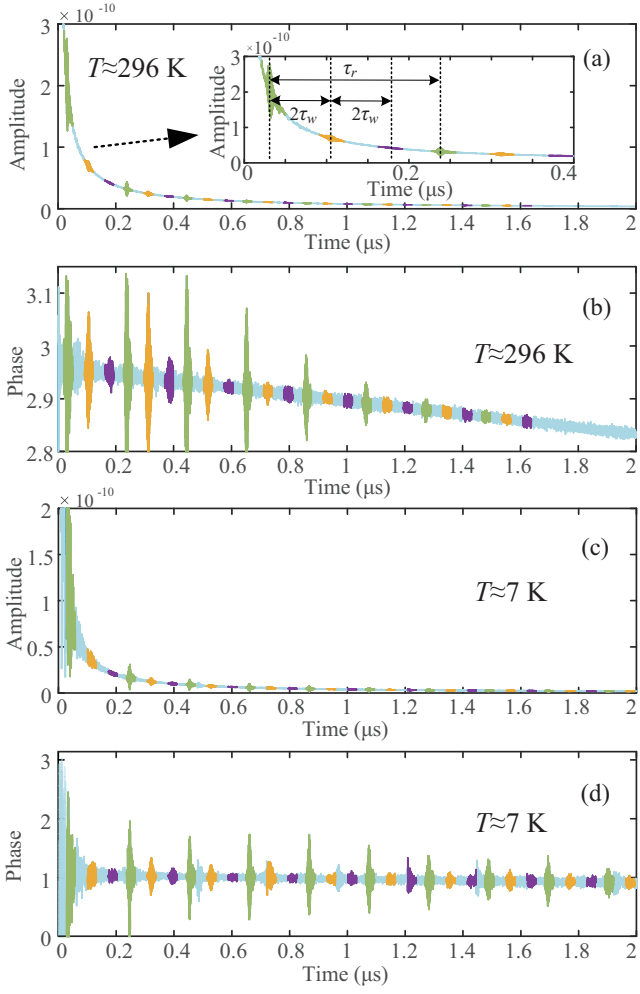}
\par\end{centering}
\caption{\label{fig:3}(a-b) The amplitude and phase of the transmission in
time domain at room temperature after a Fourier transform of the transmission spectra. The inset shows the enlarged graph
of the curve with a time scale of $0-0.4\,\mathrm{\mu s}$. (c-d) The
amplitude and phase of the transmission in time domain at $T\approx7\,\mathrm{K}$.}
\end{figure}

To eliminate the unwanted crosstalk and get an accurate $Q$-factor
of the phonon cavity, a post-processing of the spectrum is implemented. By inverse Fourier transformation, the frequency-domain transmission of the system at $0-8\,\mathrm{GHz}$
is converted to the time-domain impulse response signal of the system, as shown by the blue lines in Fig.~\ref{fig:3}.
We could imagine that when an impulse is excited and passes through
the coupling region between the waveguide and the cavity, part of
the pulse energy is coupled into the cavity. Each time the pulse in
the cavity passes through the coupling region a small fraction of
energy will couple out the cavity. So there are periodic pulses detected
with a period of the round trip time $\tau_{r}$ of the phonon pulse in
the phonon cavity, i.e. $\tau_{r}=L_{r}/v_{g,r}$, as the green pulses
show in the inset in Fig.~\ref{fig:3}(a).

Besides, there are other two groups' signals with the same period
as the yellow and purple pulses shown. These different groups of signals
are due to the acoustic reflection by the IDTs, because the impulse
will be bounced between the IDTs many times and sequentially excite
the cavity modes each time when it passes through the ring cavity.
The measured time delay $2\tau_{w}$ between the two adjacent groups
is about $2\tau_{w}=74\,\mathrm{ns}$, as is shown in Figs.~\ref{fig:3}(a)
and (b), which exactly matches the time delay due to the pulse bouncing
back and forth between the IDTs in the waveguide, i.e. $2/\varLambda=77.2\,\mathrm{ns}$.
Because the round trip time of the cavity $\tau_{r}$ is much longer
than $2\tau_{w}$, there are several detected pulses from other groups
in each round-trip time of the cavity as the inset of Fig.~\ref{fig:3}(a)
shows. Similar temporal responses of the phononic cavity are tested
for the results at $7\,\mathrm{K}$, as shown in Fig.~\ref{fig:3}(c)
and (d). The time-delayed pulses are also observed, however,
the background noise is larger compared to that at room temperature,
especially during the $0-0.2\,\mathrm{\mu s}$. The noise may be attributed to the extra electronic noise for the measurements at the cryogenic temperatures.

\begin{figure}
\begin{centering}
\includegraphics[width=1\columnwidth,height=1.1\columnwidth]{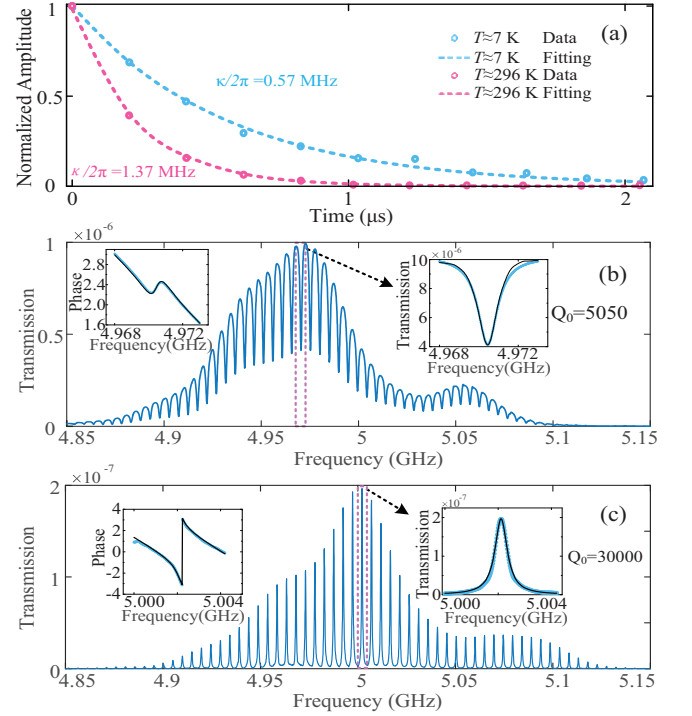}
\par\end{centering}
\caption{\label{fig:4} (a) The energy decay rate of the phonon cavity at room
temperature and $T\approx7\,\mathrm{K}$. (b-c) The transmission of the
cavity after the time domain filtering at room temperature and $T\approx7\,\mathrm{K}$.
As the first energy packet output from the cavity when  $T\approx7\,\mathrm{K}$
is neglected, the transmission shown in (c) are peaks. The insets
show the fitting of the amplitude and phase of the resonance.}
\end{figure}

To get a clean transmission spectrum of the system, we post-processe
the temporal response of the system by filtering out the green pulses,
and thus the influences due to the reflections by IDTs are mitigated.
First of all, the pulse energy is extracted and plotted in Fig.~\ref{fig:4}(a),
where the blue and red circles are the results at $T\approx7\,\mathrm{K}$
and room temperature, respectively. The evolutions of the pulse energy at these two temperatures are fitted by exponential decaying curves, giving the decay rates of $\kappa/2\pi=1.37\,\mathrm{MHz}$ and $0.57\mathrm{\,MHz}$ with the corresponding loaded $Q$-factors of $Q=3.65\times10^{3}$ and
$Q=8.772\times10^{3}$, respectively. Therefore, operating the phononic
microring at the cryogenic temperature improves the lifetime by 2.4 times.

The transmission spectra of the system are reconstructed by converting the filtered time-domain signals back to the frequency-domain using Fourier transform, and
the results are plotted in Figs.~\ref{fig:4}(b) and (c). Fig.~\ref{fig:4}(b)
shows clean resonance spectrum with negligible background noise, and
we could identify periodic resonance dips with $\mathrm{FSR}/2\pi=4.8\,\mathrm{MHz}$.
The insets show the typical results of the fitted resonance dip and
the phase, giving a loaded quality factor $Q=4.14\times10^{3}$ and
intrinsic quality factor $Q_{0}=5.05\times10^{3}$. Similarly, the
reconstructed spectrum at $T\approx7\,\mathrm{K}$ {[}Fig.~\ref{fig:4}(c){]}
shows periodic spectral features without background noise. However, the resulted Lorentz peaks instead of dips are in sharp contrast to Fig.\ref{fig:4}(b). The
reason is that when filtrating the impulses in Figs.~\ref{fig:3}(c) and
(d), the first green pulse is neglected due to the significant noise
background, and the remaining pulses correspond to the emission of
the cavity instead of the transmission of the cavity. Therefore, the
reconstructed spectrum is equivalent to a cavity emission spectrum
received from the drop port of the cavity. By fitting the amplitude and
phase of a resonance, as the insets show, we obtain the loaded
and intrinsic quality factors $Q=8.932\times10^{3}$ and $Q_{0}=3\times10^{4}$,
respectively. Comparing the results at room temperature and $7\,\mathrm{K}$,
the corresponding propagation losses of the phonon waveguide are reduced
from $\alpha\approx4.343\cdot\omega/(v_{g}\cdot Q_{i})=4.41\,\mathrm{dB\,mm^{-1}}$
to $1.29\,\mathrm{dB\,mm^{-1}}$. The intrinsic $Q$-factor has 5
times increment with the coupling between the cavity and the phonon
waveguide changing from undercoupling to overcoupling as the phase
curves shown in the insets of Figs.~\ref{fig:4} (b) and (c). Besides,
the peak of the IDT bandwidth has a blue-shift about $30\,\mathrm{MHz}$
due to the cooling down of the chip and a $fQ$ product of $1.5\times10^{14}$
is obtained at $T\approx7\,\mathrm{K}$. It should be noted that both the loaded $Q$-factors obtained by fitting the resonance bandwidth are close to the values obtained by fitting the lifetime of the phonon in
Fig.~\ref{fig:4} (a), and both the amplitude and phase can be fitted
well by the Lorentzian-type transmission of the micro-resonator.
These results, therefore, justify the time-domain filtration operations in the data processing.

\section{CONCLUSION}

In conclusion, a high frequency traveling wave phononic microring
resonator based on the high-acoustic-index-contrast mechanism is demonstrated
on GaN-on-Sapphire chip. The frequency of the phononic modes is up
to the typical working frequency of superconducting qubits at about $5\,\mathrm{GHz}$.
The phonon losses of the waveguide measured at room temperature and
 $T\approx7\,\mathrm{K}$ are $\alpha\approx4.41\,\mathrm{dB\,mm^{-1}}$
($Q_{0}=5.05\times10^{3}$) and $1.29\,\mathrm{dB\,mm^{-1}}$ ($Q_{0}=3\times10^{4}$),
respectively. A largest $fQ$ product about $1.5\times10^{14}$ is
obtained finally. The phonon loss is expected to continue decreasing
by further lowering the temperature and improving the nanofabrication
technologies. We expect that such a phononic cavity can be employed
to replace the superconducting microwave resonant cavity in the future
for circuit quantum acoustodynamics. Besides, the phononic
cavities system can be developed in the future for applications in
topological phononics and acoustic filters. By applying electric fields
in the system, phonon can be better controlled and modulated.

\begin{acknowledgments}
This work was supported by the National Natural Science Foundation of China (Grant No.12061131011, No.92165209, No.11874342, No.11922411,  No. 11925404), China Postdoctoral Science Foundation (BX2021167), Key-Area Research and Development Program of Guangdong Provice (Grant No. 2020B0303030001), the Natural Science Foundation of Anhui Provincial (Grant No. 2108085MA17 and 2108085MA22), and Grant No. 2019GQG1024 from the Institute for Guo Qiang, Tsinghua University. CLZ was also supported by the Fundamental Research Funds for the Central Universities (Grant No.WK2470000031). This work was partially carried out at the USTC Center for Micro and Nanoscale Research and Fabrication.
\end{acknowledgments}

\end{document}